\documentclass[a4paper,fleqn,usenatbib]{mnras}

\usepackage{newtxtext,newtxmath}
\usepackage[T1]{fontenc}
\usepackage{ae,aecompl}

\usepackage{graphicx}	% Including figure files
\usepackage{amsmath}	% Advanced maths commands
\title[Atomic data and electron density in PNe]{Atomic data and the density structures of planetary nebulae}

\author[Juan de Dios \& Rodr\'iguez]{ Leticia Juan de Dios\thanks{E-mail: leticiajd@inaoep.mx}
and M\'onica Rodr\'iguez,
\\
Instituto Nacional de Astrof\'isica \'Optica y Electr\'onica, Luis Enrique Erro 1, Tonantzintla 72840, Puebla, Mexico\\
}

\date{Accepted XXX. Received YYY; in original form ZZZ}

\pubyear{2015}

\begin{document}
\label{firstpage}
\pagerange{\pageref{firstpage}--\pageref{lastpage}}
\maketitle

\begin{abstract}
We study the density structures of planetary nebulae implied by four diagnostics that sample different regions within the nebulae: [\ion{S}{ii}] $\lambda6716/\lambda6731$, [\ion{O}{ii}] $\lambda3726/\lambda3729$, [\ion{Cl}{iii}] $\lambda5518/\lambda5538$, and [\ion{Ar}{iv}] $\lambda4711/\lambda4740$. We use a sample of 46 objects with deep spectra that allow the calculation of the electron density from these four diagnostics, and explore the impact that different atomic data have on the results. We compare the observational results with those obtained from photoionization models characterized by three different density structures. We conclude that the atomic data used in the calculations of electron density fully determine the density structures that are derived for the objects. We illustrate this by selecting three combinations of atomic data that lead to observational results that are compatible with each of the three different density structures explored with the models.
\end{abstract}


%%%%%%%%%%%%%%%%%%%%%%%%%%%%%%%%%%%%%%%%%%%%%%%%%%

%%%%%%%%%%%%%%%%% BODY OF PAPER %%%%%%%%%%%%%%%%%%

\section{Introduction}
Reliable chemical abundances of planetary nebulae (PNe) can provide fundamental information on the evolution of low and intermediate mass stars and on how they enrich the interstellar medium of galaxies \citep[see, e.g.,][]{DelIng15}. For these abundances to be reliable, it is critical to obtain good determinations of the physical conditions, the electron density and electron temperature, which can be estimated using the relative intensities of collisionally excited lines. The electron density is especially important at the high densities found in many PNe, where the use of different density diagnostics or different atomic data introduces significant changes in the derived chemical abundances \citep{DelIng09, JuandeDios2017}. 

In the optical range, the most used density diagnostics are [\ion{S}{ii}] $\lambda6716/\lambda6731$, [\ion{O}{ii}] $\lambda3726/\lambda3729$, [\ion{Cl}{iii}] $\lambda5518/\lambda5538$, and [\ion{Ar}{iv}] $\lambda4711/\lambda4740$. These diagnostics provide information on the density structure in the objects, since they are produced by ions that sample different regions of the nebulae. That PNe are not necessarily homogeneous can be appreciated in the multiple images that have been obtained of these objects, where knots and condensations are usually observed. Thus, the discrepancies between the electron densities obtained from different diagnostics can be indicating the presence of density variations inside PNe. However, the interpretation of the results is not straightforward. One must take into account that the lines used in the diagnostics have different critical densities \citep{Rubin89}, and this can introduce a bias in the results. Lines with relatively low critical densities, as the ones used by the [\ion{S}{ii}] and [\ion{O}{ii}] diagnostics, are suppressed in high density regions by collisional de-excitation. Thus, in a highly stratified nebula containing dense ionized clumps, the emission of the lines used in these diagnostics will be biased towards regions of low density. Besides, observational errors and uncertainties in the atomic data used in the calculations will have different effects on the results, depending on the density regime and on the density diagnostic.

Several works, such as those of \cite{Saraph1970, Stanghellini1989, Copetti2002}, and \cite{Wang04}, have studied the density structure of PNe using the four density diagnostics listed above, finding a variety of results. The four studies find that the [\ion{S}{ii}] densities are higher than those derived from [\ion{O}{ii}], and most of the authors conclude that there is good agreement between the densities obtained with [\ion{S}{ii}] and [\ion{Cl}{iii}], but \cite{Stanghellini1989} find that the electron densities obtained with [\ion{Cl}{iii}] are  higher than the ones obtained with [\ion{S}{ii}]. \cite{Saraph1970} and \cite{Wang04} find that the densities obtained for [\ion{Ar}{iv}] are usually higher than the ones implied by the other diagnostics, but \cite{Copetti2002} argue that [\ion{Cl}{iii}] and [\ion{Ar}{iv}] lead to similar results, and \cite{Stanghellini1989} find that the density obtained with [\ion{Ar}{iv}] is lower than the ones obtained with the other diagnostics. This last result was used by \cite{Stanghellini1989} to suggest that the Ar$^{+3}$ spectral lines might be produced in regions whose densities are lower than those where the other ions are located.

Since the observational samples used in these studies are large, the different results they find about the density stratification cannot be attributed to observational errors, but should be due to differences in the methodology, in particular the atomic data, adopted in each work. In fact, \cite{Copetti2002} and \cite{Wang04} argue that the systematically lower densities obtained with [\ion{O}{ii}] arise from uncertainties in the transition probabilities used for [\ion{O}{ii}].  \cite{Wang04} explore the effect of changing the atomic data for O$^{+} $ and find a set of transition probabilities that lead to a better agreement in the densities of [\ion{O}{ii}] and [\ion{S}{ii}]. However, the effects of using different atomic data have been barely explored, and might explain most of the discrepancies and the differences between the different studies \citep{JuandeDios2017}.

In \cite{JuandeDios2017}, we explored the uncertainties introduced by different atomic data in nebular abundance determinations using different combinations of the available atomic data to calculate physical conditions and chemical abundances in a sample of 36 PNe and eight \ion{H}{ii} regions. We found that atomic data introduce uncertainties that can exceed 0.6--0.8~dex at densities above $10^4${cm$^{-3}$ and that the atomic data related to the determination of electron density are the ones that have the largest impact. Previous works, such as those listed above, have mentioned the bias that the selection of atomic data introduce in the determination of the electron density, but a systematic analysis of how the different atomic datasets affect the density structure determined for PNe has not been made.

In this work, we compare the electron densities estimated from different ions using photoionization models and observations. By assuming different density structures in the models, we determine the type of patterns that can be expected for the densities derived from the [\ion{S}{ii}], [\ion{O}{ii}], [\ion{Cl}{iii}], and [\ion{Ar}{iv}] diagnostics. We also explore the effects of introducing `observational errors' in the line ratios predicted by the models. We compare the results from the models with those obtained for a sample of PNe using different combinations of atomic data. The purpose is to explore what we can know about the density structures of PNe and about the atomic data used to determine these structures.

\section{Observational Sample}
\label{sec:obs} 

Our sample consists of 46 Galactic PNe with published deep spectra of relatively high spectral resolution. The objects in the sample have the lines required to define the four most used density diagnostics, [\ion{S}{ii}] $\lambda6716/\lambda6731$, [\ion{O}{ii}] $\lambda3726/\lambda3729$, [\ion{Cl}{iii}] $\lambda5518/\lambda5538$, and [\ion{Ar}{iv}] $\lambda4711/\lambda4740$, and the two most used temperature diagnostics, [\ion{N}{ii}] $\lambda5755/(\lambda6548+\lambda6583)$ and [\ion{O}{iii}] $\lambda4363/(\lambda4959+\lambda5007)$. The objects cover a wide range of degrees of excitation: there are PNe where \ion{He}{ii} lines are not detected along with objects where the intensities of \ion{He}{ii}~$\lambda4686$ and H$\beta$ are very similar. On the other hand, the electron densities are in the range $\log(n_{\mathrm{e}})\simeq2.0$--$4.5$~cm$^{-3}$. Hence the sample can be considered representative of the population of ionized nebulae, or at least of those objects where the lines listed above are not too weak to be measured. 

The 46 PNe and the references for their spectra are: IC~2165 \citep{Hyung94}, NGC~2440 \citep{HA1998}, NGC~6153 \citep{Liu00}, IC~4846 \citep{Hyung01}, IC~5217 \citep{Hyung01b}, M~1-42, M~2-36 \citep{Liu01}, Hu~1-2 \citep{PHABBF03}, IC~418 \citep{Sharpee03}, NGC~6302 \citep{TBLDS2003}, Hu~1-2, NGC~6210, NGC~6572, NGC~6720, NGC~6741, NGC~6826, NGC~6884, NGC~7662 \citep{Liu04}, NGC~6543 \citep{WL2004}, IC~2003, NGC~6803 \citep{Wesson05}, Cn~2-1, H~1-50, He~2-118, M~1-20, M~2-4, M~3-21, M~3-32, NGC~6439, NGC~6567, NGC~6620 \citep{Wang07}, DdDm1 \citep{OHLIT09}, NGC~2867 \citep{GarciaR09}, NGC~7009 \citep{Fang11}, Hb~4, Cn~1-5, He~2-86, M~1-61, M~3-15, NGC~5189, NGC~6369, PC~14, Pe~1-1 \citep{GarciaR12}, H4-1 \citep{OT2013}, NGC~6210 \citep{BERD2015}, and H~1-50 \citep{GDGD18}.

\section{Models}
\label{sec:models} 

We have constructed a grid of spherical models using the photoionization code {\sc Cloudy} \citep[version c17.00,][]{Cloudy}. The input chemical abundances and dust properties are the ones that come by default in {\sc Cloudy}, identified as `PN abundances' and `ISM dust', respectively. The parameters that vary for each model are the effective temperature,  $T_{\mathrm{eff}}$, the luminosity of the ionizing source, $L$, the inner nebular radius, $R_0$, and the density of hydrogen in the gas, $n_{\mathrm{H}}$, as a function of the distance to the star, $R$. 

The models are ionized by blackbody spectra characterized by $T_{\mathrm{eff}}=30$, 35, 40, 50, 75, 100, 125, 150, 180, 210, $240\times10^3$~K and $L=200$, 1000, 3000, 6000, 10000, 15000~L$_{\odot}$. The distance from the central star to the gas is given by $R_0=5\times10^{15}$, $10^{16}$, $5\times10^{16}$, $3\times10^{17}$, $3\times10^{18}$, and $10^{19}$~cm. The hydrogen density at $R_0$ is in the range $10^2$--$10^5$~cm$^{-3}$ and varies in steps of 0.2 dex.

The density structures we are considering are: constant hydrogen density, a power law ($n_\mathrm{H}\propto R^\alpha$, where $\alpha$ varies from $-1.5$ to 4 in steps of 0.5), and combined structures. All the models are ionization bounded.

The combined models are constructed by adding the line intensities relative to H${\beta}$ of the spectra of two constant-density models. With this approach, each model contributes 50 per cent to the total intensity of H${\beta}$. The combined models have the same effective temperature and luminosity for the ionizing star but have different hydrogen densities and can have different initial radii. There is a large number of possible combinations of models, 26700, and we just calculate a random sample of 5000. The maximum contrast in density between the two combined models is a factor of ten. Some of the computed models, and those that we initially calculated for higher density contrasts, lead to densities from the [\ion{Cl}{iii}] diagnostic that are more than $0.2$~dex higher than the density from [\ion{O}{ii}] at $n_{\mathrm{e}}\leq5600$~cm$^{-3}$, a result that is not found in our sample PNe for any combination of atomic data. Hence, these models are not included in the final sample.

For the power-law models, some combinations of the parameters do not return a feasible model. For example, combinations of a high density at $R_0$ with a low effective temperature. If the initial hydrogen density is too high or if it increases extremely fast, there are not enough photons to ionize the gas and the models fail. Those failed models are excluded from the analysis that follows.

We have changed the transition probabilities and collision strengths that come by default in {\sc Cloudy} for the ions used to determine the physical conditions.\footnote{This was done by creating new stout format datafiles that contain the atomic data of interest (\url{https://www.nublado.org/wiki/StoutData}).} The models have been computed using two sets of atomic data, A and B. Further details are provided below in Section~\ref{sec:analysis}.

We use {\sc pyCloudy} \citep{CM13} to extract for each model the line intensities needed to use the four density diagnostics and the two temperature diagnostics mentioned in Section~\ref{sec:obs} above: $n_{\mathrm{e}}$[\ion{S}{ii}], $n_{\mathrm{e}}$[\ion{O}{ii}], $n_{\mathrm{e}}$[\ion{Cl}{iii}], $n_{\mathrm{e}}$[\ion{Ar}{iv}], $T_{\mathrm{e}}$[\ion{N}{ii}], and $T_{\mathrm{e}}$[\ion{O}{iii}]. We only take into account models where the intensities of the diagnostic lines relative to H$\beta$ are equal or higher than the faintest ones measured in the objects. With this restriction, around 50--60 per cent of the models of each type of density structure are discarded. The most critical lines to filter the models in this way are the [\ion{Ar}{iv}] lines. 

In some cases, the density diagnostics are near the limits of their validity, and roundoff errors lead to line ratios that do not return a density value. These models are also rejected from our analysis.

Initial calculations are performed for 4050, 1000, and 5000 models of constant, power-law, and combined-density structure. After all the rejection processes described above, we are left with 833 (dataset A) and 662 (dataset B) models of constant hydrogen density, 387 (dataset A) and 341 (B) power-law models, and 1260 (A) and 1648 (B) combined models. We note that the exact number of models is not very important for our purposes, since we are just illustrating the kind of density structures that can be obtained with these assumed density structures.

\section{Analysis and atomic data}
\label{sec:analysis} 

We calculate the electron densities and temperatures for both observations and models using {\sc Pyneb} (v1.1.1), a code for the analysis of nebular emission lines developed by \cite{Pyneb}. {\sc Pyneb} solves the equations of statistical equilibrium for the lower-lying energy levels of the ions we are considering here \citep[see, e.g.,][]{Osterbrock}. These equations require the use of transition probabilities and collision strengths that are obtained from quantum mechanical calculations \citep[see, e.g.,][]{Aggarwal13}.

Since we want to compare the results obtained from the observations with those implied by the models, for each model we extract the values of the line intensities necessary to obtain the electron density and electron temperature, and the analysis of these lines is performed in the same way for both observations and model.
We calculate for each object and model $n_{\mathrm{e}}$[\ion{S}{ii}], $n_{\mathrm{e}}$[\ion{O}{ii}], $n_{\mathrm{e}}$[\ion{Cl}{iii}] and $n_{\mathrm{e}}$[\ion{Ar}{iv}]. We also derive an average density from the logarithmic values of these densities, $\langle\log(n_{\mathrm{e}})\rangle$, and use this average density to determine $T_{\mathrm{e}}$[\ion{N}{ii}] and $T_{\mathrm{e}}$[\ion{O}{iii}]. We use $T_{\mathrm{e}}$[\ion{N}{ii}] to calculate $n_{\mathrm{e}}$[\ion{S}{ii}] and $n_{\mathrm{e}}$[\ion{O}{ii}], and $T_{\mathrm{e}}$[\ion{O}{iii}] for $n_{\mathrm{e}}$[\ion{Cl}{iii}] and $n_{\mathrm{e}}$[\ion{Ar}{iv}].

We start by considering the atomic data shown in Table~\ref{tab:atdata}. This database is a compilation of transition probabilities and collision strengths for the ions that we use in our analysis, N$^+$, O$^+$, O$^{++}$, S$^+$, Cl$^{++}$, and Ar$^{3+}$. The effects of most of these atomic datasets on the derived physical conditions have been previously explored in \cite{JuandeDios2017}, but we have added sets of transition probabilities for [\ion{S}{ii}] \citep{KKFBL14}, [\ion{Ar}{iv}] \citep{CK63}, and [\ion{O}{iii}] \citep{TFF01}, which are the ones used by default in {\sc CLOUDY}, a set of collision strenghts for [\ion{Cl}{iii}] \citep{ST12}, and the recent determinations of transition probabilities of \cite{Rynkun19} for [\ion{S}{ii}], [\ion{Cl}{iii}] and [\ion{Ar}{iv}]. All these sets of atomic data are available in recent versions of Pyneb \citep{Morisset2020}.

\begin{table}
\begin{center}
	\caption{Atomic data used in the analysis. The atomic datasets used in the computation and analysis of the models are identified with letters `A' and `B'. Letter `R'  identifies the datasets that have been rejected.}
	\label{tab:atdata}
	\resizebox{1\columnwidth}{!}{%
	\begin{tabular}{l l l l l}
\hline
Ion		&	$A_{ij}$ &	Set &	$\Upsilon_{ij}$ & Set	\\
\hline
N$^{+}$ 	&	GMZ97-WFD96	&		&	T11	&		\\
	           &	FFT04	&		&	HB04	&		\\
	           &	NR79-WFD96	&		&	LB94	&		\\
O$^{+}$ 	&	Z82-WFD96	&	B	&	P06-T07	& 		\\
				&	FFT04	&	A	&	Kal09	&		\\
				&	WFD96	&	R	&	P76-McLB93-v1&	A, B	\\
				&		&		&	P76-McLB93-v2	&		\\
				&		&		&	T07	&		\\
O$^{++}$ 	&	SZ00-WFD96	&		&	AK99	&		\\
					&	FFT04	&		&	LB94	&		\\
					&	GMZ97-WFD96	&		&	Pal12-AK99	&	R	\\
					&	TFF01	&		&	SSB14	&		\\
S$^{+}$ 	&	PKW09	&	A	&	TZ10	&	A, B	\\
				&	TZ10-PKW09	&	R	&	RBS96	&		\\
				&	VVF96-KHOC93	&		&		&		\\
				&	VVF96-MZ82a	&	B	&		&		\\
				&	KKFLB14	&	R	&		&		\\
				&	RGJ19	&		&		&		\\
Cl$^{++}$ 	&	M83-KS86	&		&	BZ89	&	A, B	\\
				&	Fal99	&	B	&	M83	&		\\
				&	M83	&	A	&	ST12	&		\\
				&	RGJ19	&		&	RBK01	&		\\
Ar$^{3+}$ 	&	MZ82a	&	A	&	RB97	&	A, B	\\
				&	MZ82a-KS86	&	B	&	M83	&	R	\\
				&	CK63	&	R	&	ZBL87	&		\\
				&	RGJ19	&		&		&		\\
\hline
\end{tabular}
}
\end{center}
References for the atomic data: AK99: \citet{AK99}, BZ89: \citet{BZ89}, CK63: \citet{CK63}, Fal99: \citet{Fal99}, FFT04: \citet{FFT04}, GMZ97: \citet{GMZ97}, HB04: \citet{HB04}, Kal09: \citet{Kal09}, KHOC93: \citet{KHOC93}, KKFLB14: \citet{KKFLB14}, KS86: \citet{KS86}, LB94: \citet{LB94}, M83: \citet{M83}, McLB93: \citet{McLB93}, MZ82a: \citet{MZ82a}, MZ82b: \citet{MZ82b}, NR79: \citet{NR79}, P06: \citet{P06}, P76: \citet{P76}, Pal12: \citet{Pal12}, PKW09: \citet{PKW09}, RB97: \citet{RB97}, RBK01: \citet{RBK01}, RBS96: \citet{RBS96}, RGJ19: \citet{Rynkun19}, SSB14: \citet{SSB14}, ST12: \citet{ST12}, SZ00: \citet{SZ00}, T07: \citet{T07}, T11: \citet{T11}, TZ10: \citet{TZ10}, VVF96: \citet{VVF96}, WFD96: \citet{WFD96}, Z82: \citet{Z82}, ZBL87: \citet{ZBL87}.
\end{table}

In \cite{JuandeDios2017}, we argue that four of the atomic datasets listed in Table~\ref{tab:atdata} should not be used. Three of them lead to densities that differ significantly from those inferred from the other density diagnostics in most objects, or that fail to return a valid electron density for several objects. These data are the transition probabilities of \cite{TZ10} for S$^+$, the transition probabilities of \cite{WFD96} for O$^+$, and the collision strengths of \cite{M83} for Ar$^{+3}$. The fourth dataset to be excluded contains the collision strengths of \cite{Pal12} for O$^{+3}$, whose values could not be reproduced by \cite{SSB14}. These four datasets are not used in any of the calculations that follow, and are classified as rejected and identified with an `R' in Table~\ref{tab:atdata}.

We have studied the behaviour of the added atomic datasets listed above before including them in further parts of the analysis. We have compared the results obtained using these new sets with those implied by all other datasets for the sample of PNe. We find that for 20 of the 46 objects in the sample, we cannot obtain a value of $n_{\mathrm{e}}$[\ion{S}{ii}] using the transition probabilities of \citet{KKFLB14}. On the other hand, some of the values of $n_{\mathrm{e}}$[\ion{Ar}{iv}] implied by the transition probabilities of \citet{CK63} differ by more than 0.5~dex from the ones obtained for $n_{\mathrm{e}}$[\ion{S}{ii}], $n_{\mathrm{e}}$[\ion{O}{ii}] and $n_{\mathrm{e}}$[\ion{Cl}{iii}]. The values of $n_{\mathrm{e}}$[\ion{Ar}{iv}] obtained with this set of atomic data tend to be very high, and higher than the ones obtained using the set of collision strengths of \cite{M83}, which has been rejected for leading to very high densities. Since we think that these transition probabilities for S$^{+}$ and Ar$^{+3}$ are not providing reliable results, we place them in the list of rejected atomic data and we will not use them in the analysis that follows. We have identified with the letter `R' these and the previously rejected atomic datasets in Table~\ref{tab:atdata}. 

In the case of the observed objects, the calculations of physical conditions are performed using all combinations of those sets of atomic data that have not been rejected. The total number of combinations is 622,608. In the case of the models, we use two sets of atomic data, marked as `A' and `B'  in Table \ref{tab:atdata}, to explore how the derived density structures change with atomic data. These sets have the same collision strengths but different transition probabilities, since the latter are the most critical atomic data in the determination of the electron density.

Besides, for each type of model we use Monte Carlo simulations to estimate the effects of 3 and 5 per cent Gaussian errors in the line ratios used to determine the physical conditions. These uncertainties are similar to those generally reported in the literature for these line ratios. We use 500 calculations per model since the distributions of the results do not change significantly beyond that number.

 \section{Results}
\label{sec:results} 

Fig.~\ref{fig:models} shows the resulting density structures for the models. In each panel, the differences between the electron density implied by each diagnostic and the average density, $\langle\log(n_{\mathrm{e}})\rangle$ (the average of the logarithmic values of $n_{\mathrm{e}}$[\ion{S}{ii}], $n_{\mathrm{e}}$[\ion{O}{ii}], $n_{\mathrm{e}}$[\ion{Cl}{iii}] and $n_{\mathrm{e}}$[\ion{Ar}{iv}]), are plotted as a function of the average density for the models with constant, power-law and combined hydrogen densities. The upper and middle panels of Fig.~\ref{fig:models} show the results calculated with atomic datasets A and B, respectively. These results do not include uncertainties in the model-predicted line intensity ratios. Fig.~\ref{fig:models} illustrates that the density structures derived for the models computed with atomic datasets A and B are very similar, as expected. The small differences that can be seen are due to the different sensitivity introduced by atomic data to either roundoff errors in the line intensities or small changes in temperature. The lower panels of Fig.~\ref{fig:models} show the median values of the density differences derived with each diagnostic as a function of the average density for the models calculated with atomic dataset B. The median differences are calculated in the density intervals defined by the shaded and unshaded areas in Fig.~\ref{fig:models}. 

\begin{figure*}
\includegraphics[width=1\textwidth]{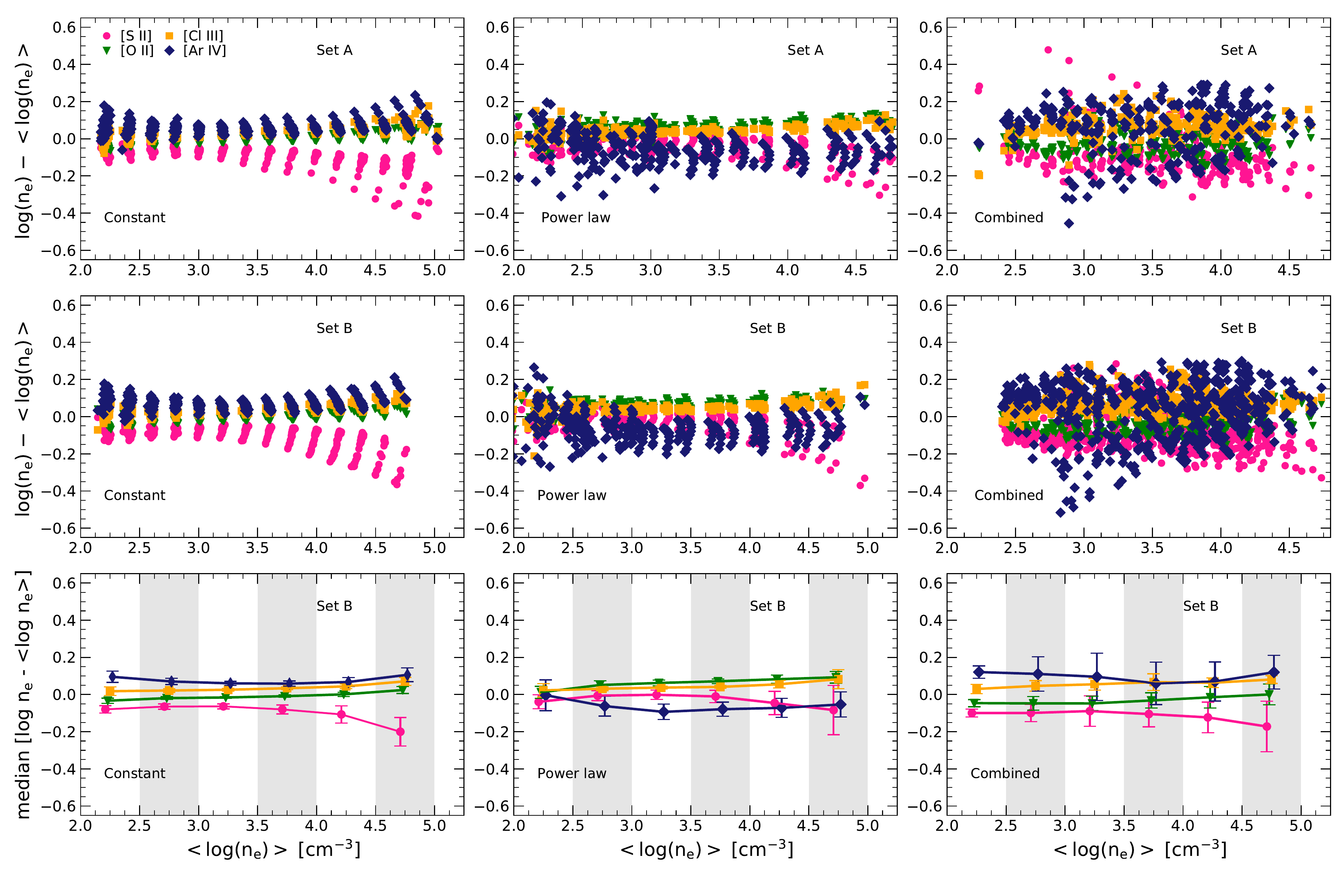}
\vspace{-0.8cm}
\caption{Differences between the density obtained from each diagnostic and the average density plotted against the average density for models with different density structures and that have been calculated using atomic datasets A (upper panels) and B (middle panels). The lower panels show the median differences for atomic dataset B, calculated for the density ranges shown by the shaded and unshaded regions. The error bars in these lower panels indicate the mean deviations, and the symbols have been displaced slightly in the horizontal direction to avoid overlapping.}
     \label{fig:models}   
\end{figure*}

The left panels of Fig.~\ref{fig:models} show that in the models of constant hydrogen density, where the ionization introduces a gradient in electron density, the densities implied by the different diagnostics are ordered according to the ionization potential of each ion, from S$^{+}$, to O$^{+}$, Cl$^{++}$ and Ar$^{+3}$.  The combined models show similar results but with more dispersion, which is introduced by combinations of models with very different degrees of ionization because of their different densities and distances to the ionizing source. In fact, the combined models can lead to cases where most of the [\ion{Ar}{iv}] emission arises from the lower density region, making the [\ion{Ar}{iv}] diagnostic the one that implies the lowest density. We note that the number of combined models calculated for sets A and B is the same, but the number of displayed models differs because more of the models analysed with set A fail to return a valid density for the four diagnostics and are excluded from the plots.

On the other hand, the density structure displayed by the power-law models is different to the ones obtained with the other models. The results plotted in Fig.~\ref{fig:models} show that the densities implied by the [\ion{Ar}{iv}]  diagnostic in the power-law models are usually lower than those obtained with [\ion{O}{ii}] and [\ion{Cl}{iii}], and similar to those obtained with [\ion{S}{ii}]. Thus the observed structure for these models is: $n_{\mathrm{e}}$[\ion{S}{ii}] $\simeq$ $n_{\mathrm{e}}$[\ion{Ar}{iv}] $<$ $n_{\mathrm{e}}$[\ion{O}{ii}] $\simeq$ $n_{\mathrm{e}}$[\ion{Cl}{iii}]. The different patterns introduced by the different assumed structures can thus be compared with those derived for the observational sample, and this comparison might provide some insight into the density structure of the observed objects.

The density structures derived for the sample of PNe using atomic datasets A and B are shown in Fig.~\ref{fig:objects}, where it can be seen that the two datasets lead to very different density structures for the observed objects. This effect is especially clear in the case of the [\ion{O}{ii}] diagnostic, which leads to some of the highest densities when atomic dataset A is used, and to some of the lowest densities when dataset B is used. Besides, none of the structures depicted in Fig.~\ref{fig:objects} look like the ones obtained with the models. The observational results are affected by uncertainties, but we can compare them with the results calculated for the models by assuming that the predicted line ratios have uncertainties. This is shown in Fig.~\ref{fig:models_err} for the models calculated with atomic dataset B and uncertainties equal to 5 per cent. A comparison of Figs.~\ref{fig:models} and \ref{fig:models_err} shows that the uncertainties increase the spread in the results but do not change in a significant way the structures obtained with the models.

\begin{figure*}
\begin{center}
	\includegraphics[width=0.7\textwidth]{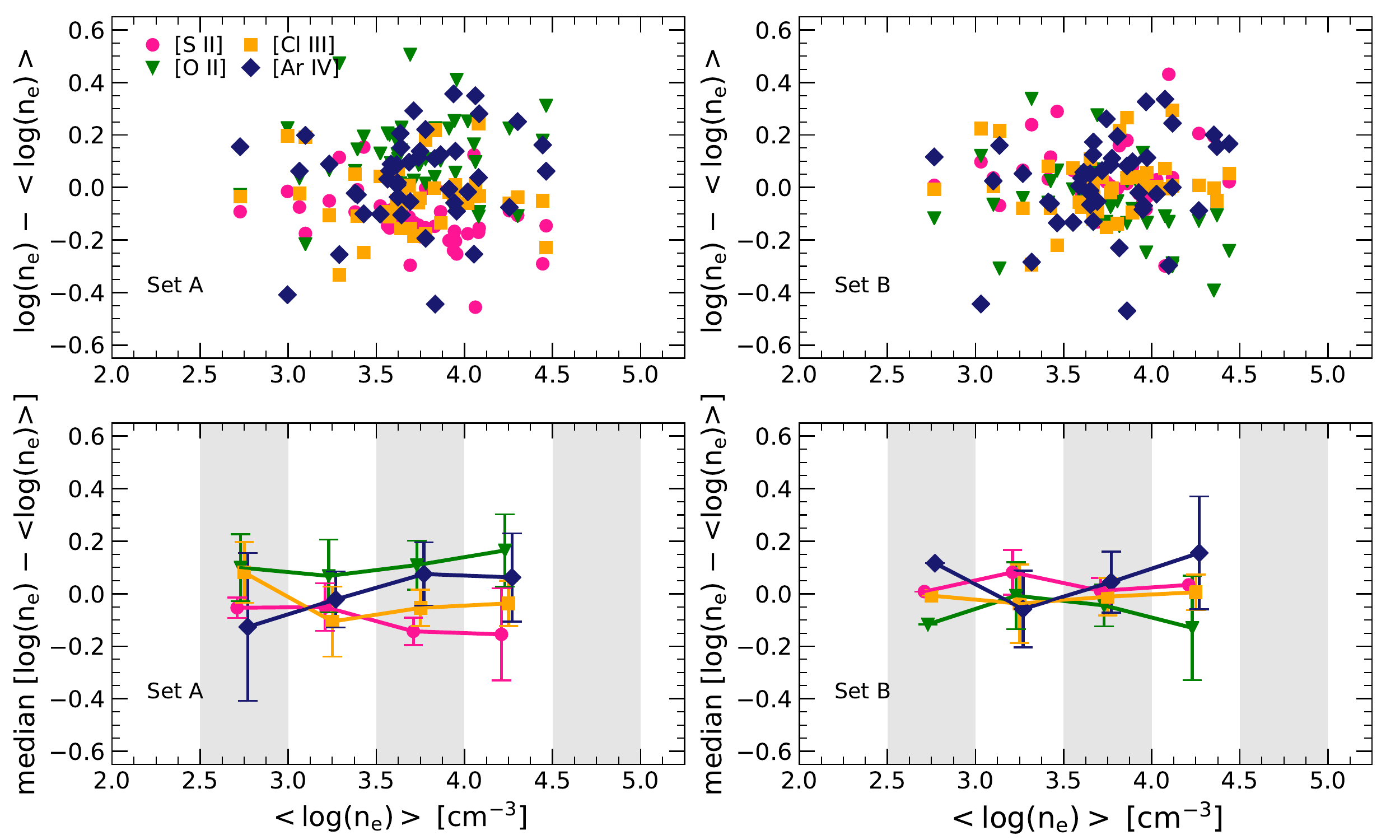}
	\vspace{-0.cm}
    \caption{Differences between the density obtained from each diagnostic and the average density as a function of the average density for the observational sample. The results are calculated using atomic datasets A (left panels) and B (right panels). The lower panels show the median differences for the density intervals defined by the shaded and unshaded areas. The error bars show the mean deviations, and the symbols have been shifted to avoid overlapping.}
    \label{fig:objects}
\end{center}
\end{figure*}

\begin{figure*}
	\includegraphics[width=1\textwidth]{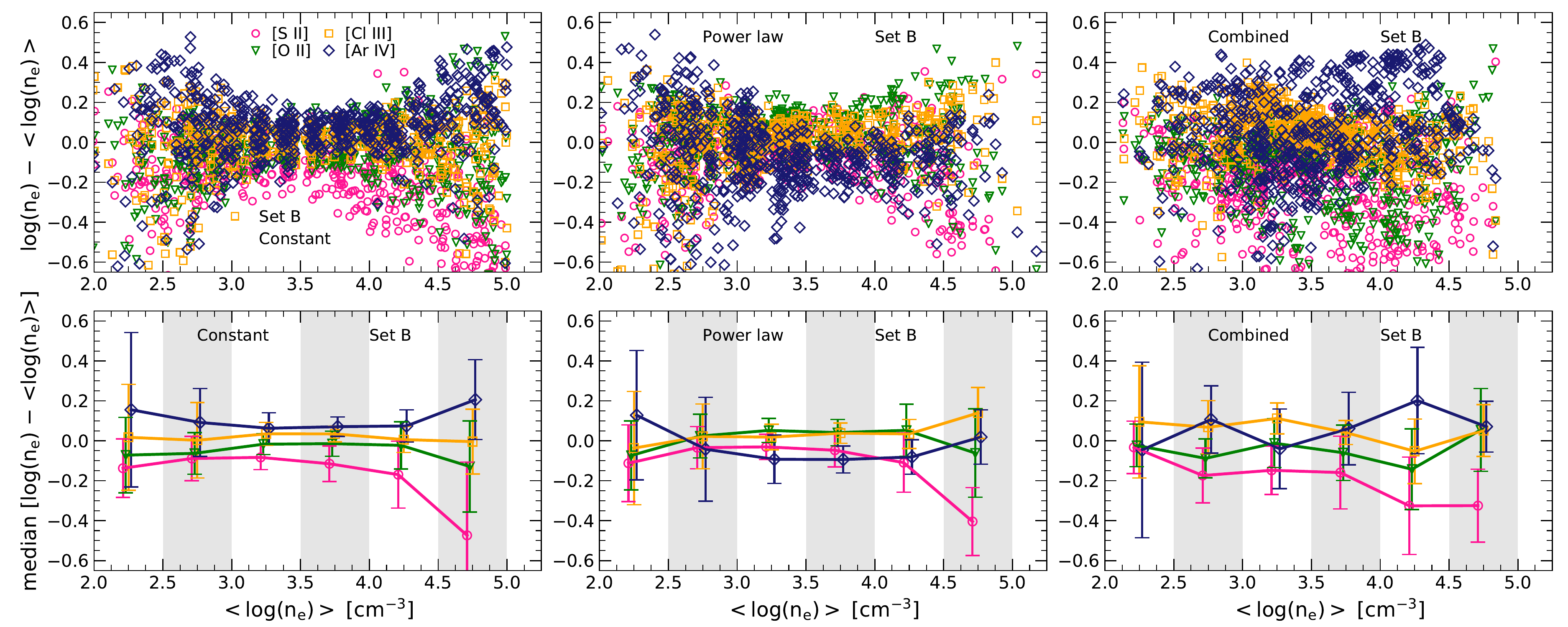}
	\vspace{-0.5cm}
    \caption{Differences between the densities obtained from each diagnostic and the average density plotted against the average density for each type of model calculated with set B. The results are a random selection of a Monte Carlo simulation that propagates errors assuming that the diagnostic line ratios of the density and temperature diagnostics have uncertainties of 5 per cent. The error bars in the lower panels show the mean deviations of the median differences, and the symbols have been shifted to avoid overlapping.}
     \label{fig:models_err}   
\end{figure*}

As a matter of fact, when other sets of atomic data are used to calculate densities for the PNe, they change the derived density structure. This means that the combination of atomic data used for the calculations determines what density structure we obtain. In order to illustrate this, we have interchanged the atomic datasets used to perform the analysis for the models of constant hydrogen density, so that the models computed with set A are analysed with set B and vice versa. The results are shown in Fig.~\ref{fig:changeAB}, where it can be seen that the change of atomic data in the analysis of the models leads to density structures similar to those derived for the objects using the same atomic data. The structure observed in the left panels of Fig.~\ref{fig:changeAB} is similar to the one obtained with dataset A for the objects, whereas the results of the right panels are similar to those obtained with dataset B for the objects. Thus, the density structures derived for the objects depend strongly on the atomic data used in the calculations: atomic data are shaping the derived density pattern.

\begin{figure*}
\begin{center}
	\includegraphics[width=0.7\textwidth]{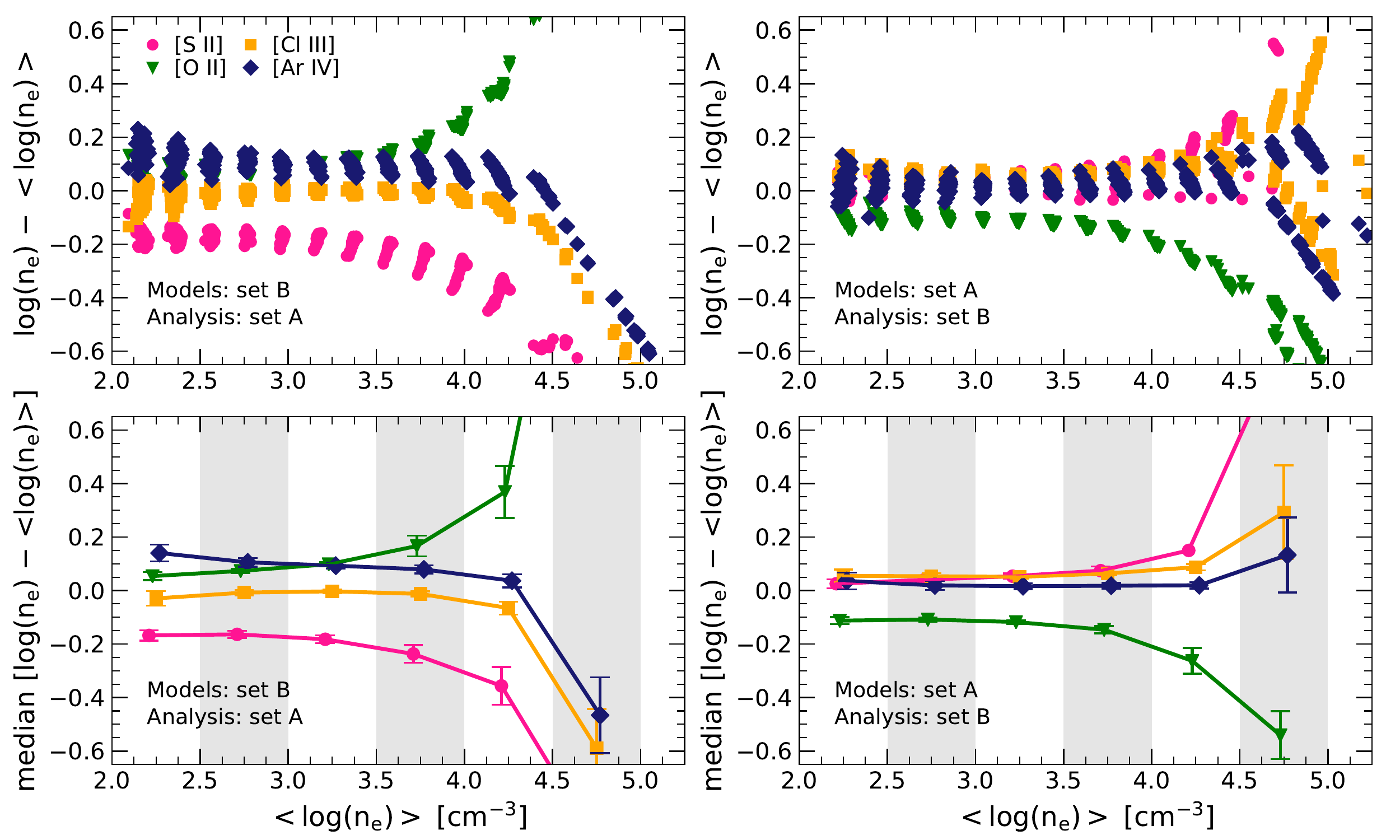}
	\vspace{-0.cm}
    \caption{Differences between the density obtained from each diagnostic and the average density as a function of the average density for models with constant density structure. The left panels show models computed with atomic dataset B and analysed with set A; the right panels show models computed with atomic dataset A and analysed with set B. The error bars in the lower panels show the mean deviations of the median differences, and the symbols have been shifted to avoid overlapping.}
    \label{fig:changeAB}   
\end{center}    
\end{figure*}

\subsection{Associating atomic datasets and density structures}\label{sec:mvso} 
 
Since the choice of atomic data determines the density structure derived for the observed objects, and since not all density structures are possible, one might wonder if we can constrain the structure of the real objects or determine which atomic data are working better by comparing the results implied by realistic structures with those derived for the observational sample using different combinations of atomic data.

The density structures we have used here to construct the photoionization models are not comprehensive and might be unrealistic, but we can use them to explore this matter. In order to do so, we want to compare the patterns observed in the models with the results obtained with different atomic data for the objects. We can see in Fig.~\ref{fig:models} that the densities obtained with [\ion{O}{ii}] and [\ion{Cl}{iii}] lead to similar values for models with different density structures. On the other hand, the [\ion{Ar}{iv}] and [\ion{S}{ii}] diagnostics are distributed in different ways around the positions defined by [\ion{O}{ii}] and [\ion{Cl}{iii}] for different structures. We can use these results to look for combinations of atomic data that reproduce these different patterns in the objects. 

Hence, for each set of observational results calculated with one specific combination of atomic data, we calculate: the median value of the difference $\log(n_\mathrm{e}$[\ion{Cl}{iii}]$)-\log(n_\mathrm{e}$[\ion{O}{ii}]); the median of the difference between $\log(n_\mathrm{e}$[\ion{Ar}{iv}]) and the average density of chlorine and oxygen, $\log(n_\mathrm{e}$[\ion{Ar}{iv}]$)-\langle\log(n_\mathrm{e}$[\ion{Cl}{iii}]), $\log(n_\mathrm{e}$[\ion{O}{ii}]$)\rangle$; and the median of the difference between $\log(n_\mathrm{e}$[\ion{S}{ii}]) and the average density of chlorine and oxygen, $\log(n_\mathrm{e}$[\ion{S}{ii}]$)-\langle\log(n_\mathrm{e}$[\ion{Cl}{iii}]), $\log(n_\mathrm{e}$[\ion{O}{ii}]$)\rangle$. These quantities are also calculated for all the model results based on atomic datasets A or B, including the results with no `observational uncertainties' along with the results that consider variations introduced by 3 and 5 per cent uncertainties in the line ratios. Fig.~\ref{fig:hist} shows the distributions of differences obtained for the objects in the sample using all the combinations of atomic data, with the vertical lines representing the values obtained for each type of density structure in the models.

\begin{figure*}
\begin{center}
	\includegraphics[width=0.9\textwidth]{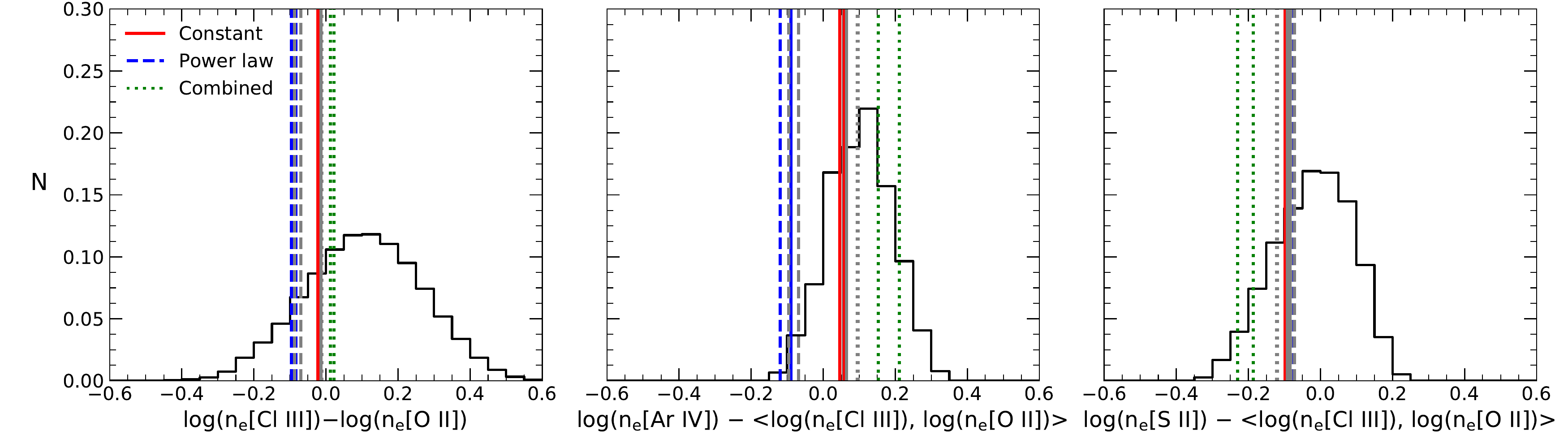}
	\vspace{-0.2cm}
    \caption{Distributions of the median values of several differences between density diagnostics. The histograms show the values obtained for the observational sample with different combinations of atomic data; the vertical lines show the values for models with constant hydrogen density (continuous lines), a power-law density structure (dashed lines), and combined structures (dotted lines). The different vertical lines include the results calculated with atomic datasets A and B and those calculated assuming uncertainties of 3 per cent in the line ratios (in red, blue, or green in the online version of this figure) and 5 per cent (in gray).}
    \label{fig:hist}   
\end{center}
\end{figure*}

The first thing that one might notice in Fig.~\ref{fig:hist} is that the lines representing the median values of the models are all in regions covered by the histograms of the observational results, which are based on different combinations of atomic data. Note that all regions covered by the histograms are in principle equally valid: if the model predictions are located at the tail of the distribution, this only means that a low number of combinations lead to a result similar to that found in the models, it does not mean that the density structure of these models is inadequate or unlikely. The question is then if there are combinations of atomic data that cover the three regions in Fig.~\ref{fig:hist} defined by each density structure in the models.

In order to answer this question, we consider the median values of $\log(n_\mathrm{e}$[\ion{Cl}{iii}]$)-\log(n_\mathrm{e}$[\ion{O}{ii}]), shown in the first panel of Fig.~\ref{fig:hist}, and find all combinations of atomic data that lead to results in the range of $\log(n_\mathrm{e}$[\ion{Cl}{iii}]$)-\log(n_\mathrm{e}$[\ion{O}{ii}]) covered by all the median values associated with each modeled density structure. This includes the model results calculated with both atomic datasets A and B (whose range of variation can provide some indication of the effect of changing the atomic data) and with and without uncertainties of 3 and 5 per cent in the diagnostic line ratios. The three sets of combinations are then required to cover the results shown in the other two panels of Fig.~\ref{fig:hist}. We find combinations of atomic data that satisfy all the restrictions for each type of density structure.  

In the case of the power law models, there is a single combination of transition probabilities and collision strengths for the ions used to determine electron density that satisfy the different criteria. On the other hand, in the case of the constant density and combined models, there are several combinations of atomic data that fall into the regions limited by each type of density structure. We have selected a combination of atomic datasets from the filtered combinations and plot the resulting density structures obtained for the objects in Fig.~\ref{fig:selected}. Table~\ref{tab:atdata_select} shows the sets of transition probabilities and collision strengths used for each type of model. We see that the selected combinations of atomic data roughly reproduce the structure derived in the models. This is especially noticeable for the atomic data that lead to density structures similar to the ones implied by the combined models.  Hence, the results of Fig.~\ref{fig:hist} imply that, given the available atomic data, none of the density structures explored with the models can be ruled out.

\begin{figure*}
\begin{center}
	\includegraphics[width=\textwidth]{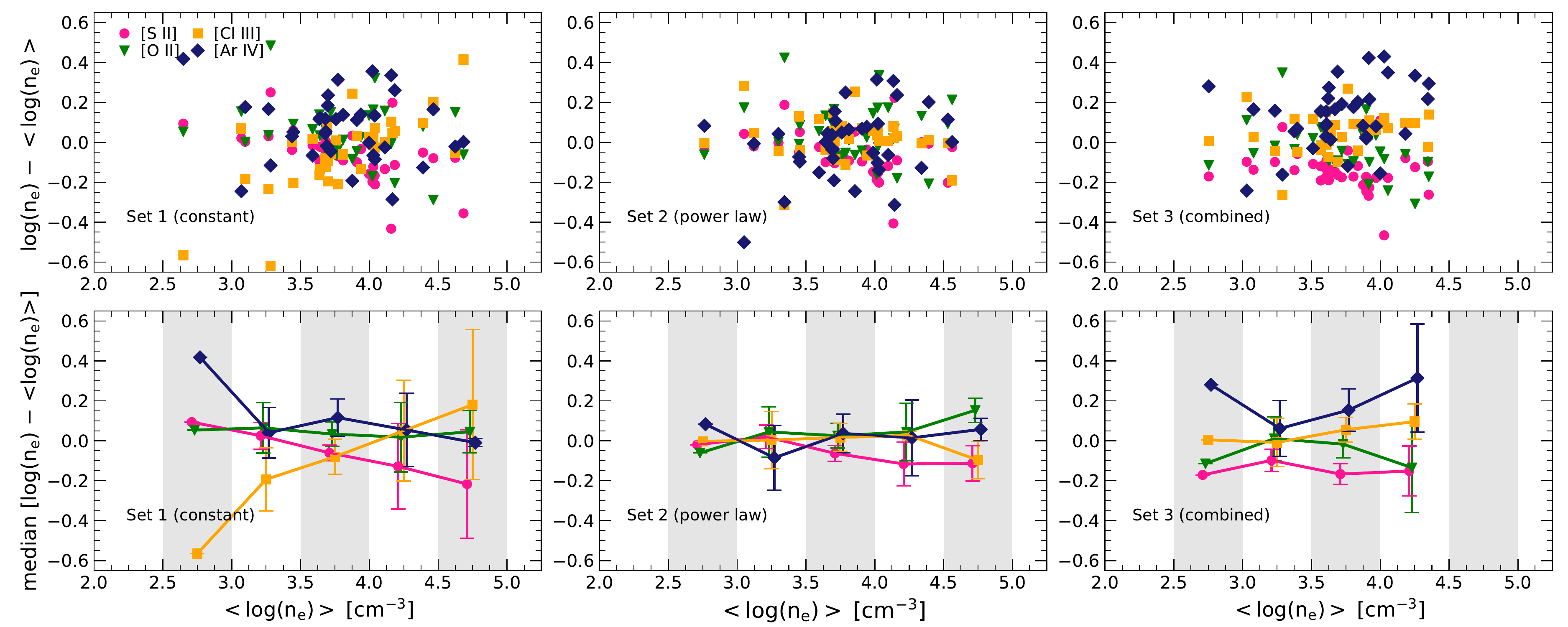}
	\vspace{-0.5cm}
    \caption{Differences between the density diagnostics and the average density plotted against the average density for the sampled objects. The results are calculated with the atomic data presented in Table \ref{tab:atdata_select}, which have been selected to reproduce the density structures explored with the models. These results can be compared with those presented in Figs.~\ref{fig:models} and \ref{fig:models_err}. The error bars in the lower panels show the mean deviations of the median differences, and the symbols have been shifted in the horizontal direction to avoid overlapping.}
   \label{fig:selected} 
\end{center}     
\end{figure*}

\begin{table*}
	\caption{Examples of atomic data that reproduce each type of density structure. The references are defined in Table~\ref{tab:atdata}.}
	\label{tab:atdata_select}
	\begin{tabular}{l l l l l l l}
\hline
		& \multicolumn{2}{l}{Set 1 (constant)}& \multicolumn{2}{l}{Set 2 (power law)}& \multicolumn{2}{l}{Set 3 (combined)}\\\\
Ion		&	$A_{ij}$ &	$\Upsilon_{ij}$ &	$A_{ij}$ & $\Upsilon_{ij}$ &	$A_{ij}$ & $\Upsilon_{ij}$\\
\hline
N$^{+}$ 	&	FFT04	&	T11	&	FFT04	&	HB04	&	FFT04	&	HB04	\\
O$^{+}$ 	&	FFT04	&	P76-McLB93-v1	&	FFT04	&	P76-McLB93-v2	&	Z82-WFD96 &	Kal09	\\
O$^{++}$ &	FFT04	&	AK99	&	SZ00-WFD96	&	SSB14	&	SZ00-WFD96	&	SSB14	\\
S$^{+}$ 	&	VVF96-MZ82a	&	RBS96	&	RGJ19	&	RBS96	&	PKW09	&	RBS96	\\
Cl$^{++}$ &	Fal99	&	M83	&	RGJ19	&	M83	&	Fal99	&	BZ89	\\
Ar$^{3+}$ &	MZ82a-KS86	&	RB97	&	RGJ19	&	RB97 &	MZ82a		&	ZBL87	\\
\hline
\end{tabular}
\end{table*}

We note that given one density structure, we can only choose the combinations of atomic data that lead to a similar structure, not the individual datasets corresponding to the collision strengths or transition probabilities of each ion. Besides, each single set of atomic data is distributed along most of the range covered by the histograms in Fig.~\ref{fig:hist}, and we cannot reject any set by considering that it lies outside any of the regions that are covered by the models. This is specially true for the atomic data used to determine the electron temperature, since all of these sets are evenly distributed across the histograms in Fig.~\ref{fig:hist}. Therefore, given the density structures explored here, no other set of atomic data can be rejected with this analysis.

PNe could have density structures more complex than the ones we have explored here, arising from all the physical processes that take place during their formation and evolution. Besides, different objects could have different structures. The density structures we are considering in this work might be too simplistic representations of the possible structures that can be present in these objects. Still, we consider them useful for exploring the effects of atomic data. We have seen that atomic data have a big impact on the results, shaping the density structures that are derived. Besides, we have shown that by choosing different combinations of the available atomic data we can obtain in principle any of the density structures studied here.

\section{Summary and conclusions}

We have explored the density structures of PNe by constructing a grid of CLOUDY photoionization models that have different hydrogen density structures: a constant density, a power-law dependence on the radius, and combined models of different constant hydrogen densities. The results are calculated with two sets of atomic data, labelled as A and B in Table~\ref{tab:atdata}, which are then used to derive the physical condition of the models using their predicted line intensity ratios and the same analysis that we use for the observational data.

The observational sample contains 46 PNe with published spectra that have measurements of the lines needed to estimate the electron density and temperature from the four most used density diagnostics, [\ion{S}{ii}] $\lambda6716/\lambda6731$, [\ion{O}{ii}] $\lambda3726/\lambda3729$, [\ion{Cl}{iii}] $\lambda5518/\lambda5538$, and [\ion{Ar}{iv}] $\lambda4711/\lambda4740$, and the two most used temperature diagnostics, [\ion{N}{ii}] $\lambda5755/(\lambda6548+\lambda6583)$ and [\ion{O}{iii}] $\lambda4363/(\lambda4959+\lambda5007)$. These spectra are analysed using all possible combinations of atomic data for these ions. 

The atomic datasets used in the calculations have been selected from those available after rejecting six datasets that lead to dicrepancies. Four of the datasets were rejected in \cite{JuandeDios2017}; and we argue that the the transition probabilities of \citet{KKFLB14} for [\ion{S}{ii}] and \citet{CK63} for [\ion{Ar}{iv}], which are used by default in CLOUDY, should also be rejected. The data of \citet{KKFLB14} cannot be used to derive $n_\mathrm{e}$[\ion{S}{ii}] in 20 of the 46 PNe in the sample, whereas the data of \citet{CK63} for [\ion{Ar}{iv}] lead to densities which are larger than the ones implied by the other diagnostics by factors above 3.

As expected, the models analysed with the same atomic dataset used to calculate the models lead to very similar results for the two data sets, A and B, for all types of density structures. However, the density structures obtained for the objects change with the atomic data and differ from the ones determined for the models with the same atomic data. If we interchange the atomic data for the analysis of the models so that the models computed with atomic dataset A are analysed with set B and vice versa, we find a large change in the density structure derived for the models, which is then similar to the structure found in the objects when they are analysed with the same set of data. This illustrates the fact that, even after filtering the most extreme datasets, the derived density structures vary significantly depending on the atomic data used: different combinations of atomic data lead to very different density structures.

We have selected three combinations of atomic data that reproduce the three density structures explored with the models. The fact that we have been able to do so indicates that we cannot determine both what density structures are predominant in PNe and which atomic data are working better. This degeneracy might be broken by considering more sophisticated models that try to reproduce a larger number of characteristics of the objects, like their temperature structure.

We have performed this analysis using a large sample of atomic data. New determinations of atomic data could be included in the future, but this will not change our main result: the density structure derived for the objects will still depend on the atomic data used in the analysis.

\section*{Data availability}
The data underlying this article will be shared on reasonable request to the corresponding author. Other datasets were obtained from sources of public domain stated in the references. 

\section*{Acknowledgements}
We thank the anonymous referee for useful comments. We also thank I.~Aretxaga and M. Richer for their useful suggestions. We acknowledge support from Mexican CONACYT grant CB-2014-240562. LJdD acknowledges support from CONACYT grant 298356.
%%%%%%%%%%%%%%%%%%%%%%%%%%%%%%%%%%%%%%%%%%%%%%%%%%


\bibliographystyle{mnras}
\bibliography{Bib}

\begin{thebibliography}{}
\makeatletter
\relax
\def\mn@urlcharsother{\let\do\@makeother \do\$\do\&\do\#\do\^\do\_\do\%\do\~}
\def\mn@doi{\begingroup\mn@urlcharsother \@ifnextchar [ {\mn@doi@}
  {\mn@doi@[]}}
\def\mn@doi@[#1]#2{\def\@tempa{#1}\ifx\@tempa\@empty \href
  {http://dx.doi.org/#2} {doi:#2}\else \href {http://dx.doi.org/#2} {#1}\fi
  \endgroup}
\def\mn@eprint#1#2{\mn@eprint@#1:#2::\@nil}
\def\mn@eprint@arXiv#1{\href {http://arxiv.org/abs/#1} {{\tt arXiv:#1}}}
\def\mn@eprint@dblp#1{\href {http://dblp.uni-trier.de/rec/bibtex/#1.xml}
  {dblp:#1}}
\def\mn@eprint@#1:#2:#3:#4\@nil{\def\@tempa {#1}\def\@tempb {#2}\def\@tempc
  {#3}\ifx \@tempc \@empty \let \@tempc \@tempb \let \@tempb \@tempa \fi \ifx
  \@tempb \@empty \def\@tempb {arXiv}\fi \@ifundefined
  {mn@eprint@\@tempb}{\@tempb:\@tempc}{\expandafter \expandafter \csname
  mn@eprint@\@tempb\endcsname \expandafter{\@tempc}}}

\bibitem[\protect\citeauthoryear{{Aggarwal} \& {Keenan}}{{Aggarwal} \&
  {Keenan}}{1999}]{AK99}
{Aggarwal} K.~M.,  {Keenan} F.~P.,  1999, \mn@doi [ApJS] {10.1086/313232},
  \href {http://adsabs.harvard.edu/abs/1999ApJS..123..311A} {123, 311}

\bibitem[\protect\citeauthoryear{{Aggarwal} \& {Keenan}}{{Aggarwal} \&
  {Keenan}}{2013}]{Aggarwal13}
{Aggarwal} K.,  {Keenan} F.,  2013, Fusion Science and Technology, \href
  {http://adsabs.harvard.edu/abs/2013arXiv1301.3002A} {63, 3}

\bibitem[\protect\citeauthoryear{{Bohigas}, {Escalante}, {Rodr{\'\i}guez}  \&
  {Dufour}}{{Bohigas} et~al.}{2015}]{BERD2015}
{Bohigas} J.,  {Escalante} V.,  {Rodr{\'\i}guez} M.,   {Dufour} R.~J.,  2015,
  \mn@doi [MNRAS] {10.1093/mnras/stu2389}, \href
  {https://ui.adsabs.harvard.edu/abs/2015MNRAS.447..817B} {447, 817}

\bibitem[\protect\citeauthoryear{{Butler} \& {Zeippen}}{{Butler} \&
  {Zeippen}}{1989}]{BZ89}
{Butler} K.,  {Zeippen} C.~J.,  1989, A\&A, \href
  {http://adsabs.harvard.edu/abs/1989A&A...208..337B} {208, 337}

\bibitem[\protect\citeauthoryear{{Copetti} \& {Writzl}}{{Copetti} \&
  {Writzl}}{2002}]{Copetti2002}
{Copetti} M.~V.~F.,  {Writzl} B.~C.,  2002, \mn@doi [A\&A]
  {10.1051/0004-6361:20011621}, \href
  {http://adsabs.harvard.edu/abs/2002A&A...382..282C} {382, 282}

\bibitem[\protect\citeauthoryear{{Czyzak} \& {Krueger}}{{Czyzak} \&
  {Krueger}}{1963}]{CK63}
{Czyzak} S.~J.,  {Krueger} T.~K.,  1963, \mn@doi [MNRAS]
  {10.1093/mnras/126.2.177}, \href
  {https://ui.adsabs.harvard.edu/abs/1963MNRAS.126..177C} {126, 177}

\bibitem[\protect\citeauthoryear{{Delgado Inglada}, {Rodr{\'\i}guez}, {Mampaso}
   \& {Viironen}}{{Delgado Inglada} et~al.}{2009}]{DelIng09}
{Delgado Inglada} G.,  {Rodr{\'\i}guez} M.,  {Mampaso} A.,   {Viironen} K.,
  2009, \mn@doi [\apj] {10.1088/0004-637X/694/2/1335}, \href
  {https://ui.adsabs.harvard.edu/abs/2009ApJ...694.1335D} {694, 1335}

\bibitem[\protect\citeauthoryear{{Delgado-Inglada}, {Rodr{\'{\i}}guez},
  {Peimbert}, {Stasi{\'n}ska}  \& {Morisset}}{{Delgado-Inglada}
  et~al.}{2015}]{DelIng15}
{Delgado-Inglada} G.,  {Rodr{\'{\i}}guez} M.,  {Peimbert} M.,  {Stasi{\'n}ska}
  G.,   {Morisset} C.,  2015, \mn@doi [MNRAS] {10.1093/mnras/stv388}, \href
  {http://adsabs.harvard.edu/abs/2015MNRAS.449.1797D} {449, 1797}

\bibitem[\protect\citeauthoryear{{Fang} \& {Liu}}{{Fang} \&
  {Liu}}{2011}]{Fang11}
{Fang} X.,  {Liu} X.-W.,  2011, \mn@doi [MNRAS]
  {10.1111/j.1365-2966.2011.18681.x}, \href
  {http://adsabs.harvard.edu/abs/2011MNRAS.415..181F} {415, 181}

\bibitem[\protect\citeauthoryear{{Ferland} et~al.,}{{Ferland}
  et~al.}{2017}]{Cloudy}
{Ferland} G.~J.,  et~al., 2017, \rmxaa, \href
  {https://ui.adsabs.harvard.edu/abs/2017RMxAA..53..385F} {53, 385}

\bibitem[\protect\citeauthoryear{Fischer \& Tachiev}{Fischer \&
  Tachiev}{2004}]{FFT04}
Fischer C.~F.,  Tachiev G.,  2004, \mn@doi [At. Data and Nucl. Data Tables]
  {http://dx.doi.org/10.1016/j.adt.2004.02.001}, 87, 1

\bibitem[\protect\citeauthoryear{{Fritzsche}, {Fricke}, {Geschke}, {Heitmann}
  \& {Sienkiewicz}}{{Fritzsche} et~al.}{1999}]{Fal99}
{Fritzsche} S.,  {Fricke} B.,  {Geschke} D.,  {Heitmann} A.,   {Sienkiewicz}
  J.~E.,  1999, \mn@doi [ApJ] {10.1086/307328}, \href
  {http://adsabs.harvard.edu/abs/1999ApJ...518..994F} {518, 994}

\bibitem[\protect\citeauthoryear{{Galavis}, {Mendoza}  \& {Zeippen}}{{Galavis}
  et~al.}{1997}]{GMZ97}
{Galavis} M.~E.,  {Mendoza} C.,   {Zeippen} C.~J.,  1997, \mn@doi [A\&AS]
  {10.1051/aas:1997344}, \href
  {http://adsabs.harvard.edu/abs/1997A&AS..123..159G} {123}

\bibitem[\protect\citeauthoryear{{Garc{\'{\i}}a-Rojas}, {Pe{\~n}a}  \&
  {Peimbert}}{{Garc{\'{\i}}a-Rojas} et~al.}{2009}]{GarciaR09}
{Garc{\'{\i}}a-Rojas} J.,  {Pe{\~n}a} M.,   {Peimbert} A.,  2009, \mn@doi
  [A\&A] {10.1051/0004-6361:200811185}, \href
  {http://adsabs.harvard.edu/abs/2009A&A...496..139G} {496, 139}

\bibitem[\protect\citeauthoryear{{Garc{\'{\i}}a-Rojas}, {Pe{\~n}a}, {Morisset},
  {Mesa-Delgado}  \& {Ruiz}}{{Garc{\'{\i}}a-Rojas} et~al.}{2012}]{GarciaR12}
{Garc{\'{\i}}a-Rojas} J.,  {Pe{\~n}a} M.,  {Morisset} C.,  {Mesa-Delgado} A.,
  {Ruiz} M.~T.,  2012, \mn@doi [A\&A] {10.1051/0004-6361/201118217}, \href
  {http://adsabs.harvard.edu/abs/2012A&A...538A..54G} {538, A54}

\bibitem[\protect\citeauthoryear{{Garc{\'\i}a-Rojas}, {Delgado-Inglada},
  {Garc{\'\i}a-Hern{\'a}ndez}, {Dell'Agli}, {Lugaro}, {Karakas}  \&
  {Rodr{\'\i}guez}}{{Garc{\'\i}a-Rojas} et~al.}{2018}]{GDGD18}
{Garc{\'\i}a-Rojas} J.,  {Delgado-Inglada} G.,  {Garc{\'\i}a-Hern{\'a}ndez}
  D.~A.,  {Dell'Agli} F.,  {Lugaro} M.,  {Karakas} A.~I.,   {Rodr{\'\i}guez}
  M.,  2018, \mn@doi [MNRAS] {10.1093/mnras/stx2519}, \href
  {https://ui.adsabs.harvard.edu/abs/2018MNRAS.473.4476G} {473, 4476}

\bibitem[\protect\citeauthoryear{{Hudson} \& {Bell}}{{Hudson} \&
  {Bell}}{2004}]{HB04}
{Hudson} C.~E.,  {Bell} K.~L.,  2004, \mn@doi [MNRAS]
  {10.1111/j.1365-2966.2004.07461.x}, \href
  {http://adsabs.harvard.edu/abs/2004MNRAS.348.1275H} {348, 1275}

\bibitem[\protect\citeauthoryear{{Hyung}}{{Hyung}}{1994}]{Hyung94}
{Hyung} S.,  1994, \mn@doi [ApJS] {10.1086/191860}, \href
  {http://adsabs.harvard.edu/abs/1994ApJS...90..119H} {90, 119}

\bibitem[\protect\citeauthoryear{{Hyung} \& {Aller}}{{Hyung} \&
  {Aller}}{1998}]{HA1998}
{Hyung} S.,  {Aller} L.~H.,  1998, \mn@doi [\pasp] {10.1086/316155}, \href
  {https://ui.adsabs.harvard.edu/abs/1998PASP..110..466H} {110, 466}

\bibitem[\protect\citeauthoryear{{Hyung}, {Aller}  \& {Lee}}{{Hyung}
  et~al.}{2001a}]{Hyung01}
{Hyung} S.,  {Aller} L.~H.,   {Lee} W.-b.,  2001a, \mn@doi [\pasp]
  {10.1086/324415}, \href {http://adsabs.harvard.edu/abs/2001PASP..113.1559H}
  {113, 1559}

\bibitem[\protect\citeauthoryear{{Hyung}, {Aller}, {Feibelman}  \&
  {Lee}}{{Hyung} et~al.}{2001b}]{Hyung01b}
{Hyung} S.,  {Aller} L.~H.,  {Feibelman} W.~A.,   {Lee} W.-B.,  2001b, \mn@doi
  [\aj] {10.1086/321171}, \href
  {http://adsabs.harvard.edu/abs/2001AJ....122..954H} {122, 954}

\bibitem[\protect\citeauthoryear{{Juan de Dios} \& {Rodr{\'{\i}}guez}}{{Juan de
  Dios} \& {Rodr{\'{\i}}guez}}{2017}]{JuandeDios2017}
{Juan de Dios} L.,  {Rodr{\'{\i}}guez} M.,  2017, \mn@doi [MNRAS]
  {10.1093/mnras/stx916}, \href
  {http://adsabs.harvard.edu/abs/2017MNRAS.469.1036J} {469, 1036}

\bibitem[\protect\citeauthoryear{{Kaufman} \& {Sugar}}{{Kaufman} \&
  {Sugar}}{1986}]{KS86}
{Kaufman} V.,  {Sugar} J.,  1986, \mn@doi [Journal of Physical and Chemical
  Reference Data] {10.1063/1.555775}, \href
  {http://adsabs.harvard.edu/abs/1986JPCRD..15..321K} {15, 321}

\bibitem[\protect\citeauthoryear{{Keenan}, {Hibbert}, {Ojha}  \&
  {Conlon}}{{Keenan} et~al.}{1993}]{KHOC93}
{Keenan} F.~P.,  {Hibbert} A.,  {Ojha} P.~C.,   {Conlon} E.~S.,  1993, \mn@doi
  [\physscr] {10.1088/0031-8949/48/2/001}, \href
  {http://adsabs.harvard.edu/abs/1993PhyS...48..129K} {48, 129}

\bibitem[\protect\citeauthoryear{{Kisielius}, {Storey}, {Ferland}  \&
  {Keenan}}{{Kisielius} et~al.}{2009}]{Kal09}
{Kisielius} R.,  {Storey} P.~J.,  {Ferland} G.~J.,   {Keenan} F.~P.,  2009,
  \mn@doi [MNRAS] {10.1111/j.1365-2966.2009.14989.x}, \href
  {http://adsabs.harvard.edu/abs/2009MNRAS.397..903K} {397, 903}

\bibitem[\protect\citeauthoryear{Kisielius, Kulkarni, Ferland, Bogdanovich  \&
  Lykins}{Kisielius et~al.}{2013}]{KKFBL14}
Kisielius R.,  Kulkarni V.~P.,  Ferland G.~J.,  Bogdanovich P.,   Lykins M.~L.,
   2013, \mn@doi [The Astrophysical Journal] {10.1088/0004-637x/780/1/76}, 780,
  76

\bibitem[\protect\citeauthoryear{{Kisielius}, {Kulkarni}, {Ferland},
  {Bogdanovich}  \& {Lykins}}{{Kisielius} et~al.}{2014}]{KKFLB14}
{Kisielius} R.,  {Kulkarni} V.~P.,  {Ferland} G.~J.,  {Bogdanovich} P.,
  {Lykins} M.~L.,  2014, \mn@doi [\apj] {10.1088/0004-637X/780/1/76}, \href
  {https://ui.adsabs.harvard.edu/abs/2014ApJ...780...76K} {780, 76}

\bibitem[\protect\citeauthoryear{{Lennon} \& {Burke}}{{Lennon} \&
  {Burke}}{1994}]{LB94}
{Lennon} D.~J.,  {Burke} V.~M.,  1994, A\&AS, \href
  {http://adsabs.harvard.edu/abs/1994A&AS..103..273L} {103}

\bibitem[\protect\citeauthoryear{{Liu}, {Storey}, {Barlow}, {Danziger}, {Cohen}
   \& {Bryce}}{{Liu} et~al.}{2000}]{Liu00}
{Liu} X.-W.,  {Storey} P.~J.,  {Barlow} M.~J.,  {Danziger} I.~J.,  {Cohen} M.,
   {Bryce} M.,  2000, \mn@doi [MNRAS] {10.1046/j.1365-8711.2000.03167.x}, \href
  {http://adsabs.harvard.edu/abs/2000MNRAS.312..585L} {312, 585}

\bibitem[\protect\citeauthoryear{{Liu}, {Luo}, {Barlow}, {Danziger}  \&
  {Storey}}{{Liu} et~al.}{2001}]{Liu01}
{Liu} X.-W.,  {Luo} S.-G.,  {Barlow} M.~J.,  {Danziger} I.~J.,   {Storey}
  P.~J.,  2001, \mn@doi [MNRAS] {10.1046/j.1365-8711.2001.04676.x}, \href
  {http://adsabs.harvard.edu/abs/2001MNRAS.327..141L} {327, 141}

\bibitem[\protect\citeauthoryear{{Liu}, {Liu}, {Luo}  \& {Barlow}}{{Liu}
  et~al.}{2004}]{Liu04}
{Liu} Y.,  {Liu} X.-W.,  {Luo} S.-G.,   {Barlow} M.~J.,  2004, \mn@doi [MNRAS]
  {10.1111/j.1365-2966.2004.08155.x}, \href
  {http://adsabs.harvard.edu/abs/2004MNRAS.353.1231L} {353, 1231}

\bibitem[\protect\citeauthoryear{{Luridiana}, {Morisset}  \&
  {Shaw}}{{Luridiana} et~al.}{2015}]{Pyneb}
{Luridiana} V.,  {Morisset} C.,   {Shaw} R.~A.,  2015, \mn@doi [A\&A]
  {10.1051/0004-6361/201323152}, \href
  {http://adsabs.harvard.edu/abs/2015A&A...573A..42L} {573, A42}

\bibitem[\protect\citeauthoryear{{McLaughlin} \& {Bell}}{{McLaughlin} \&
  {Bell}}{1993}]{McLB93}
{McLaughlin} B.~M.,  {Bell} K.~L.,  1993, \mn@doi [ApJ] {10.1086/172635}, \href
  {http://adsabs.harvard.edu/abs/1993ApJ...408..753M} {408, 753}

\bibitem[\protect\citeauthoryear{{Mendoza}}{{Mendoza}}{1983}]{M83}
{Mendoza} C.,  1983, in {Flower} D.~R.,  ed., Proc. IAU Symp. 103 Planetary
  Nebulae, p. 143.

\bibitem[\protect\citeauthoryear{{Mendoza} \& {Zeippen}}{{Mendoza} \&
  {Zeippen}}{1982a}]{MZ82a}
{Mendoza} C.,  {Zeippen} C.~J.,  1982a, \mn@doi [\mnras]
  {10.1093/mnras/198.1.127}, \href
  {http://adsabs.harvard.edu/abs/1982MNRAS.198..127M} {198, 127}

\bibitem[\protect\citeauthoryear{{Mendoza} \& {Zeippen}}{{Mendoza} \&
  {Zeippen}}{1982b}]{MZ82b}
{Mendoza} C.,  {Zeippen} C.~J.,  1982b, \mn@doi [\mnras]
  {10.1093/mnras/199.4.1025}, \href
  {http://adsabs.harvard.edu/abs/1982MNRAS.199.1025M} {199, 1025}

\bibitem[\protect\citeauthoryear{{Morisset}}{{Morisset}}{2013}]{CM13}
{Morisset} C.,  2013, {pyCloudy: Tools to manage astronomical Cloudy
  photoionization code} (\mn@eprint {ascl} {1304.020})

\bibitem[\protect\citeauthoryear{{Morisset}, {Luridiana}, {Garc{\'\i}a-Rojas},
  {G{\'o}mez-Llanos}, {Bautista}, {Mendoza}  \& {Claudio}}{{Morisset}
  et~al.}{2020}]{Morisset2020}
{Morisset} C.,  {Luridiana} V.,  {Garc{\'\i}a-Rojas} J.,  {G{\'o}mez-Llanos}
  V.,  {Bautista} M.,  {Mendoza}  {Claudio} 2020, \mn@doi [Atoms]
  {10.3390/atoms8040066}, \href
  {https://ui.adsabs.harvard.edu/abs/2020Atoms...8...66M} {8, 66}

\bibitem[\protect\citeauthoryear{{Nussbaumer} \& {Rusca}}{{Nussbaumer} \&
  {Rusca}}{1979}]{NR79}
{Nussbaumer} H.,  {Rusca} C.,  1979, A\&A, \href
  {http://adsabs.harvard.edu/abs/1979A&A....72..129N} {72, 129}

\bibitem[\protect\citeauthoryear{Osterbrock \& Ferland}{Osterbrock \&
  Ferland}{2006}]{Osterbrock}
Osterbrock D.~E.,  Ferland G.~J.,  2006, {Astrophysics of Gaseous Nebulae and
  Active Galactic Nuclei}, second edn.
 Vol. 1, University Science Brooks

\bibitem[\protect\citeauthoryear{{Otsuka} \& {Tajitsu}}{{Otsuka} \&
  {Tajitsu}}{2013}]{OT2013}
{Otsuka} M.,  {Tajitsu} A.,  2013, \mn@doi [ApJ] {10.1088/0004-637X/778/2/146},
  \href {https://ui.adsabs.harvard.edu/abs/2013ApJ...778..146O} {778, 146}

\bibitem[\protect\citeauthoryear{{Otsuka}, {Hyung}, {Lee}, {Izumiura}  \&
  {Tajitsu}}{{Otsuka} et~al.}{2009}]{OHLIT09}
{Otsuka} M.,  {Hyung} S.,  {Lee} S.-J.,  {Izumiura} H.,   {Tajitsu} A.,  2009,
  \mn@doi [ApJ] {10.1088/0004-637X/705/1/509}, \href
  {https://ui.adsabs.harvard.edu/abs/2009ApJ...705..509O} {705, 509}

\bibitem[\protect\citeauthoryear{{Palay}, {Nahar}, {Pradhan}  \&
  {Eissner}}{{Palay} et~al.}{2012}]{Pal12}
{Palay} E.,  {Nahar} S.~N.,  {Pradhan} A.~K.,   {Eissner} W.,  2012, \mn@doi
  [MNRAS] {10.1111/j.1745-3933.2012.01252.x}, \href
  {http://adsabs.harvard.edu/abs/2012MNRAS.423L..35P} {423, L35}

\bibitem[\protect\citeauthoryear{Podobedova, Kelleher  \& Wiese}{Podobedova
  et~al.}{2009}]{PKW09}
Podobedova L.~I.,  Kelleher D.~E.,   Wiese W.~L.,  2009, \mn@doi [Journal of
  Physical and Chemical Reference Data] {10.1063/1.3032939}, 38, 171

\bibitem[\protect\citeauthoryear{{Pottasch}, {Hyung}, {Aller}, {Beintema},
  {Bernard-Salas}, {Feibelman}  \& {Kl{\"o}ckner}}{{Pottasch}
  et~al.}{2003}]{PHABBF03}
{Pottasch} S.~R.,  {Hyung} S.,  {Aller} L.~H.,  {Beintema} D.~A.,
  {Bernard-Salas} J.,  {Feibelman} W.~A.,   {Kl{\"o}ckner} H.~R.,  2003,
  \mn@doi [A\&A] {10.1051/0004-6361:20030104}, \href
  {https://ui.adsabs.harvard.edu/abs/2003A&A...401..205P} {401, 205}

\bibitem[\protect\citeauthoryear{{Pradhan}}{{Pradhan}}{1976}]{P76}
{Pradhan} A.~K.,  1976, \mn@doi [MNRAS] {10.1093/mnras/177.1.31}, \href
  {http://adsabs.harvard.edu/abs/1976MNRAS.177...31P} {177, 31}

\bibitem[\protect\citeauthoryear{{Pradhan}, {Montenegro}, {Nahar}  \&
  {Eissner}}{{Pradhan} et~al.}{2006}]{P06}
{Pradhan} A.~K.,  {Montenegro} M.,  {Nahar} S.~N.,   {Eissner} W.,  2006,
  \mn@doi [MNRAS] {10.1111/j.1745-3933.2005.00119.x}, \href
  {http://adsabs.harvard.edu/abs/2006MNRAS.366L...6P} {366, L6}

\bibitem[\protect\citeauthoryear{{Ramsbottom} \& {Bell}}{{Ramsbottom} \&
  {Bell}}{1997}]{RB97}
{Ramsbottom} C.~A.,  {Bell} K.~L.,  1997, \mn@doi [At. Data and Nucl. Data
  Tables] {10.1006/adnd.1997.0741}, \href
  {http://adsabs.harvard.edu/abs/1997ADNDT..66...65R} {66, 65}

\bibitem[\protect\citeauthoryear{{Ramsbottom}, {Bell}  \&
  {Stafford}}{{Ramsbottom} et~al.}{1996}]{RBS96}
{Ramsbottom} C.~A.,  {Bell} K.~L.,   {Stafford} R.~P.,  1996, \mn@doi [At. Data
  and Nucl. Data Tables] {10.1006/adnd.1996.0009}, \href
  {http://adsabs.harvard.edu/abs/1996ADNDT..63...57R} {63, 57}

\bibitem[\protect\citeauthoryear{Ramsbottom, Bell  \& Keenan}{Ramsbottom
  et~al.}{2001}]{RBK01}
Ramsbottom C.,  Bell K.,   Keenan F.,  2001, \mn@doi [Atomic Data and Nuclear
  Data Tables] {https://doi.org/10.1006/adnd.2000.0846}, 77, 57

\bibitem[\protect\citeauthoryear{Rubin}{Rubin}{1989}]{Rubin89}
Rubin R.~H.,  1989, The Astrophysical Journal Supp. Ser., 69, 897

\bibitem[\protect\citeauthoryear{{Rynkun}, {Gaigalas}  \&
  {J{\"o}nsson}}{{Rynkun} et~al.}{2019}]{Rynkun19}
{Rynkun} P.,  {Gaigalas} G.,   {J{\"o}nsson} P.,  2019, \mn@doi [A\&A]
  {10.1051/0004-6361/201834931}, \href
  {https://ui.adsabs.harvard.edu/abs/2019A&A...623A.155R} {623, A155}

\bibitem[\protect\citeauthoryear{{Saraph} \& {Seaton}}{{Saraph} \&
  {Seaton}}{1970}]{Saraph1970}
{Saraph} H.~E.,  {Seaton} M.~J.,  1970, \mn@doi [MNRAS]
  {10.1093/mnras/148.3.367}, \href
  {http://adsabs.harvard.edu/abs/1970MNRAS.148..367S} {148, 367}

\bibitem[\protect\citeauthoryear{{Sharpee}, {Williams}, {Baldwin}  \& {van
  Hoof}}{{Sharpee} et~al.}{2003}]{Sharpee03}
{Sharpee} B.,  {Williams} R.,  {Baldwin} J.~A.,   {van Hoof} P.~A.~M.,  2003,
  \mn@doi [ApJS] {10.1086/378321}, \href
  {http://adsabs.harvard.edu/abs/2003ApJS..149..157S} {149, 157}

\bibitem[\protect\citeauthoryear{{Sossah} \& {Tayal}}{{Sossah} \&
  {Tayal}}{2012}]{ST12}
{Sossah} A.~M.,  {Tayal} S.~S.,  2012, \mn@doi [ApS]
  {10.1088/0067-0049/202/2/12}, \href
  {https://ui.adsabs.harvard.edu/abs/2012ApJS..202...12S} {202, 12}

\bibitem[\protect\citeauthoryear{{Stanghellini} \& {Kaler}}{{Stanghellini} \&
  {Kaler}}{1989}]{Stanghellini1989}
{Stanghellini} L.,  {Kaler} J.~B.,  1989, \mn@doi [ApJ] {10.1086/167751}, \href
  {http://adsabs.harvard.edu/abs/1989ApJ...343..811S} {343, 811}

\bibitem[\protect\citeauthoryear{{Storey} \& {Zeippen}}{{Storey} \&
  {Zeippen}}{2000}]{SZ00}
{Storey} P.~J.,  {Zeippen} C.~J.,  2000, \mn@doi [MNRAS]
  {10.1046/j.1365-8711.2000.03184.x}, \href
  {http://adsabs.harvard.edu/abs/2000MNRAS.312..813S} {312, 813}

\bibitem[\protect\citeauthoryear{{Storey}, {Sochi}  \& {Badnell}}{{Storey}
  et~al.}{2014}]{SSB14}
{Storey} P.~J.,  {Sochi} T.,   {Badnell} N.~R.,  2014, \mn@doi [MNRAS]
  {10.1093/mnras/stu777}, \href
  {http://adsabs.harvard.edu/abs/2014MNRAS.441.3028S} {441, 3028}

\bibitem[\protect\citeauthoryear{Tachiev \& Fischer}{Tachiev \&
  Fischer}{2001}]{TFF01}
Tachiev G.,  Fischer C.,  2001, \mn@doi [Canadian Journal of Physics]
  {10.1139/p01-059}, 79, 955

\bibitem[\protect\citeauthoryear{{Tayal}}{{Tayal}}{2007}]{T07}
{Tayal} S.~S.,  2007, \mn@doi [ApJS] {10.1086/513107}, \href
  {http://adsabs.harvard.edu/abs/2007ApJS..171..331T} {171, 331}

\bibitem[\protect\citeauthoryear{{Tayal}}{{Tayal}}{2011}]{T11}
{Tayal} S.~S.,  2011, \mn@doi [ApJS] {10.1088/0067-0049/195/2/12}, \href
  {http://adsabs.harvard.edu/abs/2011ApJS..195...12T} {195, 12}

\bibitem[\protect\citeauthoryear{{Tayal} \& {Zatsarinny}}{{Tayal} \&
  {Zatsarinny}}{2010}]{TZ10}
{Tayal} S.~S.,  {Zatsarinny} O.,  2010, \mn@doi [ApJS]
  {10.1088/0067-0049/188/1/32}, \href
  {http://adsabs.harvard.edu/abs/2010ApJS..188...32T} {188, 32}

\bibitem[\protect\citeauthoryear{{Tsamis}, {Barlow}, {Liu}, {Danziger}  \&
  {Storey}}{{Tsamis} et~al.}{2003}]{TBLDS2003}
{Tsamis} Y.~G.,  {Barlow} M.~J.,  {Liu} X.~W.,  {Danziger} I.~J.,   {Storey}
  P.~J.,  2003, \mn@doi [MNRAS] {10.1046/j.1365-8711.2003.06972.x}, \href
  {https://ui.adsabs.harvard.edu/abs/2003MNRAS.345..186T} {345, 186}

\bibitem[\protect\citeauthoryear{{Verner}, {Verner}  \& {Ferland}}{{Verner}
  et~al.}{1996}]{VVF96}
{Verner} D.~A.,  {Verner} E.~M.,   {Ferland} G.~J.,  1996, \mn@doi [At. Data
  and Nucl. Data Tables] {10.1006/adnd.1996.0018}, \href
  {http://adsabs.harvard.edu/abs/1996ADNDT..64....1V} {64, 1}

\bibitem[\protect\citeauthoryear{{Wang} \& {Liu}}{{Wang} \&
  {Liu}}{2007}]{Wang07}
{Wang} W.,  {Liu} X.-W.,  2007, \mn@doi [MNRAS]
  {10.1111/j.1365-2966.2007.12198.x}, \href
  {http://adsabs.harvard.edu/abs/2007MNRAS.381..669W} {381, 669}

\bibitem[\protect\citeauthoryear{{Wang}, {Liu}, {Zhang}  \& {Barlow}}{{Wang}
  et~al.}{2004}]{Wang04}
{Wang} W.,  {Liu} X.-W.,  {Zhang} Y.,   {Barlow} M.~J.,  2004, \mn@doi [A\&A]
  {10.1051/0004-6361:20041470}, \href
  {http://adsabs.harvard.edu/abs/2004A&A...427..873W} {427, 873}

\bibitem[\protect\citeauthoryear{{Wesson} \& {Liu}}{{Wesson} \&
  {Liu}}{2004}]{WL2004}
{Wesson} R.,  {Liu} X.~W.,  2004, \mn@doi [MNRAS]
  {10.1111/j.1365-2966.2004.07856.x}, \href
  {https://ui.adsabs.harvard.edu/abs/2004MNRAS.351.1026W} {351, 1026}

\bibitem[\protect\citeauthoryear{{Wesson}, {Liu}  \& {Barlow}}{{Wesson}
  et~al.}{2005}]{Wesson05}
{Wesson} R.,  {Liu} X.-W.,   {Barlow} M.~J.,  2005, \mn@doi [MNRAS]
  {10.1111/j.1365-2966.2005.09325.x}, \href
  {http://adsabs.harvard.edu/abs/2005MNRAS.362..424W} {362, 424}

\bibitem[\protect\citeauthoryear{Wiese, Fuhr  \& Deters}{Wiese
  et~al.}{1996}]{WFD96}
Wiese W.~L.,  Fuhr J.~R.,   Deters T.~M.,  1996, Journal of Physical and
  Chemical Reference Data, Monograph 7, 403

\bibitem[\protect\citeauthoryear{{Zeippen}}{{Zeippen}}{1982}]{Z82}
{Zeippen} C.~J.,  1982, \mn@doi [MNRAS] {10.1093/mnras/198.1.111}, \href
  {http://adsabs.harvard.edu/abs/1982MNRAS.198..111Z} {198, 111}

\bibitem[\protect\citeauthoryear{{Zeippen}, {Butler}  \& {Le
  Bourlot}}{{Zeippen} et~al.}{1987}]{ZBL87}
{Zeippen} C.~J.,  {Butler} K.,   {Le Bourlot} J.,  1987, A\&A, \href
  {http://adsabs.harvard.edu/abs/1987A&A...188..251Z} {188, 251}

\makeatother
\end{thebibliography}

%%%%%%%%%%%%%%%%%%%%%%%%%%%%%%%%%%%%%%%%%%%%%%%%%%

% Don't change these lines
\bsp	% typesetting comment
\label{lastpage}
\end{document}